\documentclass[12pt]{iopart}

\usepackage{indentfirst}
\usepackage{graphicx}
\usepackage{amsmath,amssymb}
\usepackage{xcolor} 
\usepackage{hyperref}
\usepackage{amsfonts}
\usepackage{subcaption}
\usepackage{appendix}
\begin{document}

\title[Demonstration of tilt sensing using a homodyne quadrature interferometric sensor]{Demonstration of tilt sensing using a homodyne quadrature interferometric translational sensor}

\author{K Nagano \textsuperscript{1, 2}, K Mori \textsuperscript{3}, and K Izumi \textsuperscript{1}}

\address{$^1$Institute of Space and Astronautical Science, Japan Aerospace Exploration Agency, Sagamihara, Kanagawa, 252-5210, Japan}
\address{$^2$ LQUOM, Inc., Yokohama, Kanagawa, 240-8501, Japan}
\address{$^3$ Hosei University, Koganei, Tokyo, 184-0002, Japan}
\ead{koji.nagano@lquom.com}
\vspace{10pt}
\begin{indented}
\item[]\today
\end{indented}

\begin{abstract}
Future gravitational wave observation in space will demand improvement in the sensitivity of the local sensor for the drag-free control. 
This paper presents the proposal, design, and demonstration of a new laser interferometric sensor named Quadrature Interferometric Metrology of Translation and Tilt (QUIMETT) for the drag-free local sensor. 
QUIMETT enables simultaneous measurements of both translational displacement and tilts of a reflective object with a single interferometer package. 
QUIMETT offers a characteristic feature where the sensitivity to tilt is independent of the interference condition while maintaining the ability to measure the translational displacement for a range greater than the laser wavelength. 
The tilt-sensing function has been demonstrated in a prototype experiment. 
The tilt sensitivity remained unchanged in different interference conditions and stayed at $10$~nrad/Hz$^{1/2}$ at $0.1$~Hz.
\end{abstract}

%
%
%
%
%

\section{Introduction}
Gravitational waves were directly detected for the first time by Advanced Laser Interferometer Gravitational-wave Observatory (LIGO) in 2015~\cite{aligo,GWdetected1}. 
Currently, the second half of the fourth observing run, also known as O4b, by terrestrial gravitational wave antennas including LIGO, Virgo~\cite{virgo} is being performed, and KAGRA~\cite{KAGRA} will also be taking part in the observation run. 
In the meantime, space gravitational wave antennas including LISA~\cite{lisa}, TianQin~\cite{tianqin}, Taiji~\cite{taiji}, and DECIGO~\cite{decigo} have been in development or under consideration. 
The space gravitational wave antennas will probe gravitational waves oscillating at frequencies lower than a few Hz. 
Therefore, they are complementary to the terrestrial observatories. 
The space gravitational wave antennas are anticipated to promote the further research in the field of gravitational wave astronomy and astrophysics.

The space gravitational wave antennas consist of inter-satellite laser interferometers with floating test masses aboard the satellites. 
The test masses serve as stable references for the gravitational wave observation. 
To isolate the test masses from noise associated with the random motion of the satellite, the relative distance between the test mass and satellite has to be measured with local sensors and controlled by using the satellite thrusters. 
This scheme is known as the drag-free control. 
In the drag-free control, the sensing system often employs the capacitive sensors~\cite{armano2015lisa,capacitive2} which offer a wide readout range and a decent displacement sensitivity of typically nm/Hz$^{1/2}$~\cite{armano2017capacitive} with the hardware in a highly integrated form.

For the space gravitational wave antennas seeking for a higher sensitivity, the improved sensitivity for the local sensors is necessary. 
For instance, DECIGO requires a displacement sensitivity of $10^{-12}$~m/Hz$^{1/2}$ for the local sensors~\cite{nagano2020control}. 
A promising sensor concept is the quadrature readout interferometer~\cite{quadrature} which provides the displacement measurement of the test mass for a range greater than the laser wavelength with a sensitivity better than the capacitive ones. 
In fact, there have been several research efforts developing the quadrature interferometers for precision measurement applications, such as Homodyne Quadrature Interferometer (HoQI)~\cite{hoqi,HoQI2}, and modulation-assisted sensors~\cite{heinzel2010deep, gerberding2015deep, shaddock2007digitally, sutton2012digitally}. 
In addition, various solutions have been proposed to conduct multi-degrees-of-freedom measurement using interferometric sensors for broad applications
~\cite{hsieh2015development,yan2022high,yang2020single,meshksar2020applying,Xin2024}. 

Here, we propose a new sensor based on the HoQI scheme, Quadrature Interferometric Metrology of Translation and Tilt (QUIMETT). 
QUIMETT is a sensor with a simple configuration and no radio-frequency modulation is required, and is intended for use in space due to its resource-saving advantages with multi-degrees-of-freedom measurement in a single unit. 
QUIMETT adopts a linear combination of two tilt sensor outputs where the interference effect can be compensated to make the tilt sensing independent of the interference condition. 
This scheme makes QUIMETT capable of measuring the translation and two tilt degrees of freedom simultaneously. 
We have demonstrated the tilt sensing function and confirmed that the sensitivities do not vary as a function of the interference condition as expected. 
In this paper, we describe the working principle of the tilt measurement, design of a new optical system dedicated for the demonstration, and experimental verification. 

This paper is organized as follows. 
The principle of operation of the QUIMETT is presented in Section~\ref{sec:wp}. 
The optical setup of QUIMETT is given in Section~\ref{sec:es}. 
The experimental demonstration of the tilt sensing is presented in Section~\ref{sec:re}. 
After discussing the results in Section~\ref{sec:dis}, we conclude this paper in Section~\ref{sec:con}.

\begin{figure}
    \centering
    \includegraphics[width=\columnwidth]{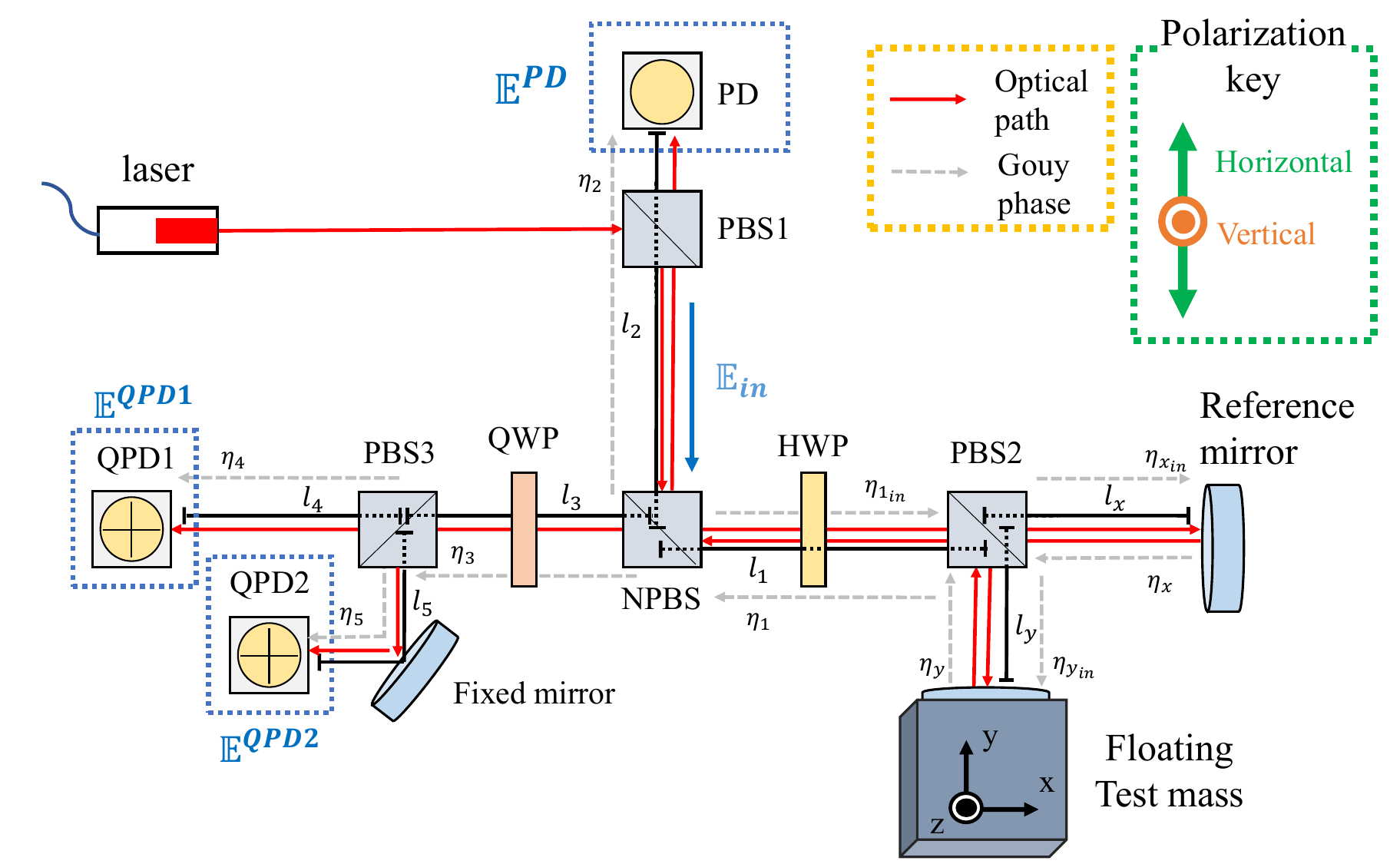}
    \caption{Conceptual setup of QUIMETT. The definition of the optical path lengths, $l$, and the Gouy phase shift along optical paths, $\eta$, are shown in Tab.~\ref{tab:l_and_eta}. PD stands for photodetector, QPD for quadrant photodetector, NPBS for non-polarizing beamsplitter, PBS for polarizing beamsplitter, HWP for half-wave plate, and QWP for quarter-wave plate.}
    \label{fig:concept}
\end{figure}

\section{Working principles \label{sec:wp}}
Figure~\ref{fig:concept} shows an illustration of the conceptual setup for QUIMETT. 
We begin the analysis by computing the electric fields. 
Subsequently, the principles of translational displacement and tilt measurements are explained. 

\subsection{Electric fields in QUIMETT}
The conventional Jones matrix formalism is extended to accommodate higher-order Hermite-Gaussian modes, so that the electric fields are handled as
\begin{equation}
    \mathbb{E}=\begin{bmatrix}
        \vec{E^h}\\\vec{E^v}
        \label{eq:Ein}
    \end{bmatrix}
\quad
\textrm{and}\quad
    \vec{E}^A=\begin{bmatrix}
        E^A_{00}\\E^A_{10}\\E^A_{01}
    \end{bmatrix},
\end{equation}
where $A$ denotes polarization content i.e., $A\in (h,v)$ with $h$ and $v$ for horizontal and vertical polarizations, respectively. 
Horizontal polarization is defined as vibrating within the $x$-$y$ plane, while vertical polarization is aligned along the $z$-axis. The electric field $E^A_{ij}$ represents the Hermite-Gauss contents for the $(i,j)$ mode. 
We hereafter denote the $6\times 6$ matrices and $6\times 1$ vectors in blackboard bold font while the $3\times 3$ matrices and $3\times 1$ vectors are written with superscripts hat and arrow, respectively.

We now derive the electric fields at the photodetectors namely, PD and QPDs 1 and 2 in Fig.~\ref{fig:concept}, by letting the laser light propagate through the optics. 
The input field $\mathbb{E}_\textrm{in}$ denotes the one departing PBS1 for non-polarizing beamsplitter (NPBS) and is set as $ \mathbb{E}_\textrm{in}= (0,0,0,\sqrt{P_\textrm{in}},0,0)^\intercal$ with $P_\textrm{in}$ the input power to the NPBS. 
Exerting the optical element matrices to the input field along the light paths, one can express the electric field vector at PD, $\mathbb{E}^\textrm{PD}$, as
\begin{equation}
\begin{split}
    \mathbb{E}^\textrm{PD}=&\mathbb{B}_h(1)\mathbb{P}(l_2,\eta_{2})\mathbb{A}_\textrm{r}\left(\frac{1}{2}\right)\mathbb{P}(l_1,\eta_{1})\mathbb{W}^{\left(2\right)}\left(-\frac{\pi}{8}\right)\\&\times\lbrace\underbrace{\mathbb{B}_v\left(R_\textrm{PBS2}\right)\mathbb{P}(l_y,\eta_\mathrm{ITF})\mathbb{M}\left(\Theta_\textrm{yaw},\Theta_\textrm{pitch}\right)\mathbb{P}(l_y,\eta_\mathrm{ITF_{in}})\mathbb{B}_v\left(R_\textrm{PBS2}\right)}_{\textrm{test mass}}\\&+\underbrace{\mathbb{B}_h(T_\textrm{PBS2})\mathbb{P}(l_x,\eta_\mathrm{ITF})\mathbb{A}_\textrm{r}(1)\mathbb{P}(l_x,\eta_\mathrm{ITF_{in}})\mathbb{B}_h\left(T_\textrm{PBS2}\right)}_{\textrm{reference mirror}}\rbrace\mathbb{W}^{\left(2\right)}\left(\frac{\pi}{8}\right)\mathbb{P}(l_1,\eta_{1_\mathrm{in}})\mathbb{A}_\textrm{r}\left(\frac{1}{2}\right)\mathbb{E}_\textrm{in},
    \label{eq:EPD}
    \end{split}
\end{equation}
where $\mathbb{A}_{r}$, $\mathbb{B}_A$, $\mathbb{W}^{(2)}$ and $\mathbb{P}$ are the optical element matrices for non-polarizing reflection, polarizing beamsplitter (PBS), half-wave plate (HWP) and free-space propagation, respectively. 
The definitions of these optical element matrices are given in~\ref{AppendixA}. 
$l$ and $\eta$ are the optical path lengths and the Gouy phase shift along optical paths, respectively. 
The detailed definitions of $l$ and $\eta$ are shown in Fig.~\ref{fig:concept} and Tab.~\ref{tab:l_and_eta}. 
Here and hereafter, we assume that $l_x$ and $l_y$ are macroscopically the same and Gouy phase shifts in these paths are the same, i.e. $\eta_{x_\mathrm{in}} = \eta_{y_\mathrm{in}} \equiv \eta_\mathrm{ITF_{in}}$ and $\eta_{x} = \eta_{y} \equiv \eta_\mathrm{ITF}$. 
$R_\textrm{PBS2}$ and $T_\textrm{PBS2}$ are reflectance for vertical polarization and transmittance for horizontal polarization of PBS2, respectively. 
The matrix $\mathbb{M}$ represents the reflection by the test mass with misalignment in both pitch and yaw directions by the normalized tilts $\Theta_\textrm{yaw}$ and $\Theta_\textrm{pitch}$~\cite{hefetz}, respectively, as
\begin{equation}
\mathbb{M}\left(\Theta_\textrm{yaw},\Theta_\textrm{pitch}\right) = \mathbb{A}_\textrm{r}(1) + \begin{bmatrix}
        \hat{M}\left(\Theta_\textrm{yaw},\Theta_\textrm{pitch}\right)&\hat{0}\\\hat{0}&\hat{M}\left(\Theta_\textrm{yaw},\Theta_\textrm{pitch}\right)
    \end{bmatrix},
\end{equation}
where $\hat{A}_r$, $\hat{M}$ and $\hat{0}$ are reflection of the mirror, misalignment and zero matrices for each polarization, respectively, the definitions of which are given in \ref{AppendixA}. 

\begin{table}[htb]
    \centering
    \caption{Definition of the optical path lengths and the Gouy phase shift along optical paths shown in Fig.~\ref{fig:concept}.}
    \label{tab:l_and_eta}
    \begin{tabular}{c|cc}
        Optical path & Length & Gouy phase shift  \\ \hline \hline
        NBPS $\rightarrow$ PBS2 & $l_1$ & $\eta_{1_\mathrm{in}}$ \\
        PBS2 $\rightarrow$ Reference mirror & $l_x$ & $\eta_{x_\mathrm{in}} (= \eta_\mathrm{ITF_{in}})$ \\
        Reference mirror $\rightarrow$ PBS2 & $l_x$ & $\eta_{x} (= \eta_\mathrm{ITF})$ \\
        PBS2 $\rightarrow$ Floating test mass & $l_y$ & 
        $\eta_{y_\mathrm{in}} (= \eta_\mathrm{ITF_{in}})$ \\
        Floating test mass $\rightarrow$ PBS2  & $l_y$ & $\eta_{y} (= \eta_\mathrm{ITF})$ \\
        PBS2 $\rightarrow$ NBPS & $l_1$ & $\eta_{1}$ \\
        NPBS $\rightarrow$ PD & $l_2$ & $\eta_{2}$ \\
        NPBS $\rightarrow$ PBS3 & $l_3$ & $\eta_{3}$ \\
        PBS3 $\rightarrow$ QPD1 & $l_4$ & $\eta_{4}$ \\
        PBS3 $\rightarrow$ QPD2 & $l_5$ & $\eta_{5}$ \\
    \end{tabular}
\end{table}

In the same fashion, the electric fields incident upon the two QPDs can be derived as
\begin{equation}
\begin{split}
 \mathbb{E}^{\textrm{QPD}n}=&\mathbb{F}_n\mathbb{W}^{\left(4\right)}\left(\frac{\pi}{4}\right)\mathbb{P}(l_3,\eta_{3})\mathbb{A}_\textrm{t}\left(\frac{1}{2}\right)\mathbb{P}(l_1,\eta_{1})\mathbb{W}^{\left(2\right)}\left(-\frac{\pi}{8}\right)\\&\times\lbrace\underbrace{\mathbb{B}_v\left(R_\textrm{PBS2}\right)\mathbb{P}(l_y,\eta_\mathrm{ITF})\mathbb{M}\left(\Theta_\textrm{yaw},\Theta_\textrm{pitch}\right)\mathbb{P}(l_y,\eta_\mathrm{ITF_{in}})\mathbb{B}_v\left(R_\textrm{PBS2}\right)}_\textrm{test mass}\\&+\underbrace{\mathbb{B}_h(T_\textrm{PBS2})\mathbb{P}(l_x,\eta_\mathrm{ITF})\mathbb{A}_\textrm{r}(1)\mathbb{P}(l_x,\eta_\mathrm{ITF_{in}})\mathbb{B}_h\left(T_\textrm{PBS2}\right)}_\textrm{reference mirror}\rbrace \mathbb{W}^{\left(2\right)}\left(\frac{\pi}{8}\right)\mathbb{P}(l_1,\eta_{1_\mathrm{in}})\mathbb{A}_\textrm{r}\left(\frac{1}{2}\right)\mathbb{E}_\textrm{in} , \label{eq:EQPD}
 \end{split}
\end{equation}
where $\mathbb{A}_\textrm{t}$ is the optical element matrices for non-polarizing transmittance, $\mathbb{W}^{(4)}$ represents a quarter-wave plate as defined in \ref{AppendixA}, and we have introduced $\mathbb{F}_n \ (n = 1, 2)$ for selecting the light path between those for QPDs 1 and 2. 
The definition is given as 
\begin{equation}
\mathbb{F}_1=\mathbb{P}(l_4,\eta_{4})\mathbb{B}_h(1),\quad\textrm{and}\quad\mathbb{F}_2=\mathbb{P}(l_5,\eta_{5})\mathbb{A}_\textrm{r}(1)\mathbb{B}_v(1).
\end{equation}
Here, we assume that the reflectance and transmittance of PBS3 are unity.
Note that proper control of the Gouy phase is crucial in our measurement approach, as it directly influences the response to tilt deviations. 

\subsection{Translational displacement readout}
From Eqs.~(\ref{eq:EPD}) and (\ref{eq:EQPD}), the laser intensity of the fundamental Gaussian mode at PD and two QPDs are derived as
\begin{align}
    I_\textrm{PD} &=\frac{1-a\cos{2k(l_x-l_y)}}{8}\frac{R_\textrm{PBS2}^2+T_\textrm{PBS2}^2}{2}P_\textrm{in}+\mathcal{O}(\Theta^2) ,
    \label{eq:IofPD} \\
    I_\textrm{QPD1} &=\frac{1-a\sin{2k(l_x-l_y)}}{8}\frac{R_\textrm{PBS2}^2+T_\textrm{PBS2}^2}{2}P_\textrm{in}+\mathcal{O}(\Theta^2) ,
    \label{eq:IofQPD1} \\
    I_\textrm{QPD2} &=\frac{1+a\sin{2k(l_x-l_y)}}{8}\frac{R_\textrm{PBS2}^2+T_\textrm{PBS2}^2}{2}P_\textrm{in}+\mathcal{O}(\Theta^2) ,
    \label{eq:IofQPD2}
\end{align}
where $\mathcal{O}(\cdot)$ is the Landau symbol, $k$ is the wave number, and $a$  denotes the visibility, defined by $a ={2R_\textrm{PBS2}T_\textrm{PBS2}}/\left(R_\textrm{PBS2}^2+T_\textrm{PBS2}^2\right)$. 
Here, we only consider the visibility degradation caused by imbalance in PBS2 while other factors are ignored, such as misalignment in HWP, spatial mode mismatching, losses in the Michelson part. 

Combining these three signals, one can read out the translational displacement for an extended range. 
For example, one can plug the measurement values into the function~\cite{hoqi},
\begin{equation}
    \frac{I_\textrm{PD}-I_\textrm{QPD2}}{I_\textrm{PD}-I_\textrm{QPD1}}=\tan{\left(2k(l_x-l_y)+\frac{\pi}{4} \right)},
    \label{eq:dispreadout}
\end{equation}
and retrieve the displacement by applying the arc-tangent operation. 
This highlights the advantage of the quadrature readout interferometer where the range of translational displacement can be expanded to a range greater than the wavelength via additional signal processing.
QUIMETT maintains this ability since $I_\textrm{QPD1}$ ($I_\textrm{QPD2}$) can be measured with the sum signal of all segments in QPD1 (QPD2).

\subsection{Tilt readout}
The QPDs play a crucial role in deriving the information associated with the tilts. 
The tilt signal for the yaw (pitch) direction can be readily obtained by taking the difference in light intensities between the left and right (upper and lower) segments. 
This results in tilt signals in the form of
    \begin{align}
       I^\textrm{tilt}_\textrm{QPD1} &= \frac{\Theta_BR_\textrm{PBS2}}{2\sqrt{2\pi}}[-T_\textrm{PBS2}\cos{\lbrace2k\left(l_x-l_y\right)}-\eta_\mathrm{ITF} -\eta_\mathrm{QPD1}\rbrace+R_\textrm{PBS2}\sin{\left(\eta_\mathrm{ITF}+\eta_\mathrm{QPD1}\right)}]P_\textrm{in},
        \label{eq:angQPD1} \\
        I^\textrm{tilt}_\textrm{QPD2} &=\frac{\Theta_BR_\textrm{PBS2}}{2\sqrt{2\pi}}[T_\textrm{PBS2}\cos{\lbrace2k\left(l_x-l_y\right)}-\eta_\mathrm{ITF}-\eta_\mathrm{QPD2}\rbrace+R_\textrm{PBS2}\sin{\left(\eta_\mathrm{ITF}+\eta_\mathrm{QPD2}\right)}]P_\textrm{in},
        \label{eq:angQPD2} \\
        \eta_\mathrm{QPD1} &\equiv \eta_1 + \eta_3 + \eta_4, \ \ \eta_\mathrm{QPD2} \equiv \eta_1 + \eta_3 + \eta_5,
    \end{align}
where $B$ is any tilt in pitch or yaw, and Eqs.~(\ref{eq:angQPD1}) and~(\ref{eq:angQPD2}) give the response to the tilt in the direction selected at $B$. 
In deriving Eqs.~(\ref{eq:angQPD1}) and~(\ref{eq:angQPD2}), we assume that the beam displacement from the center of the QPDs is small for both the reference and target beams. 
We also assume that the beam diameter is significantly smaller than the QPDs size, allowing us to neglect truncation effects. 
Additionally, we consider a setup without a lens between the target mirror and the QPDs, so the beam undergoes natural diffraction as it propagates, and its wavefront evolves without additional focusing effects.

It is now evident with Eqs.~(\ref{eq:angQPD1}) and (\ref{eq:angQPD2}) that the ordinary signal extraction with QPD is strongly influenced by the interference condition or more specifically the length difference $l_x-l_y$. 
Therefore, these signals do not directly appear to be suitable for precise tilt measurements. 

We propose introducing the sum of two tilt signals, Eqs.~(\ref{eq:angQPD1}) and (\ref{eq:angQPD2}),
\begin{align}
    I^\textrm{tilt}_\textrm{QPD1}+I^\textrm{tilt}_\textrm{QPD2} 
    &= \frac{\Theta_BR_\textrm{PBS2}^2}{2\sqrt{2\pi}}\left[\sin{\left(\eta_\mathrm{ITF}+\eta_\mathrm{QPD1}\right)} + \sin{\left(\eta_\mathrm{ITF}+\eta_\mathrm{QPD2}\right)}\right] P_\textrm{in} \\
    &= \frac{\Theta_BR_\textrm{PBS2}^2\sin{\left(\eta_\mathrm{ITF}+\eta_\mathrm{QPD}\right)}}{\sqrt{2\pi}}P_\textrm{in}.
    \label{eq:QUIMETTyaw}
\end{align}
Here and hereafter, we assume that $l_4$ and $l_5$ are the same and, consequently, $\eta_4 = \eta_5$ and $\eta_\mathrm{QPD1} = \eta_\mathrm{QPD2} \equiv \eta_\mathrm{QPD}$. 
The tilts along both measurement axes of the target can be directly determined, independent of interference effect $l_x-l_y$. 
Note that the effect of the reference mirror tilt is discussed in \ref{AppendixB}.

\begin{figure}[t]
    \centering
    \includegraphics[ width=\columnwidth]{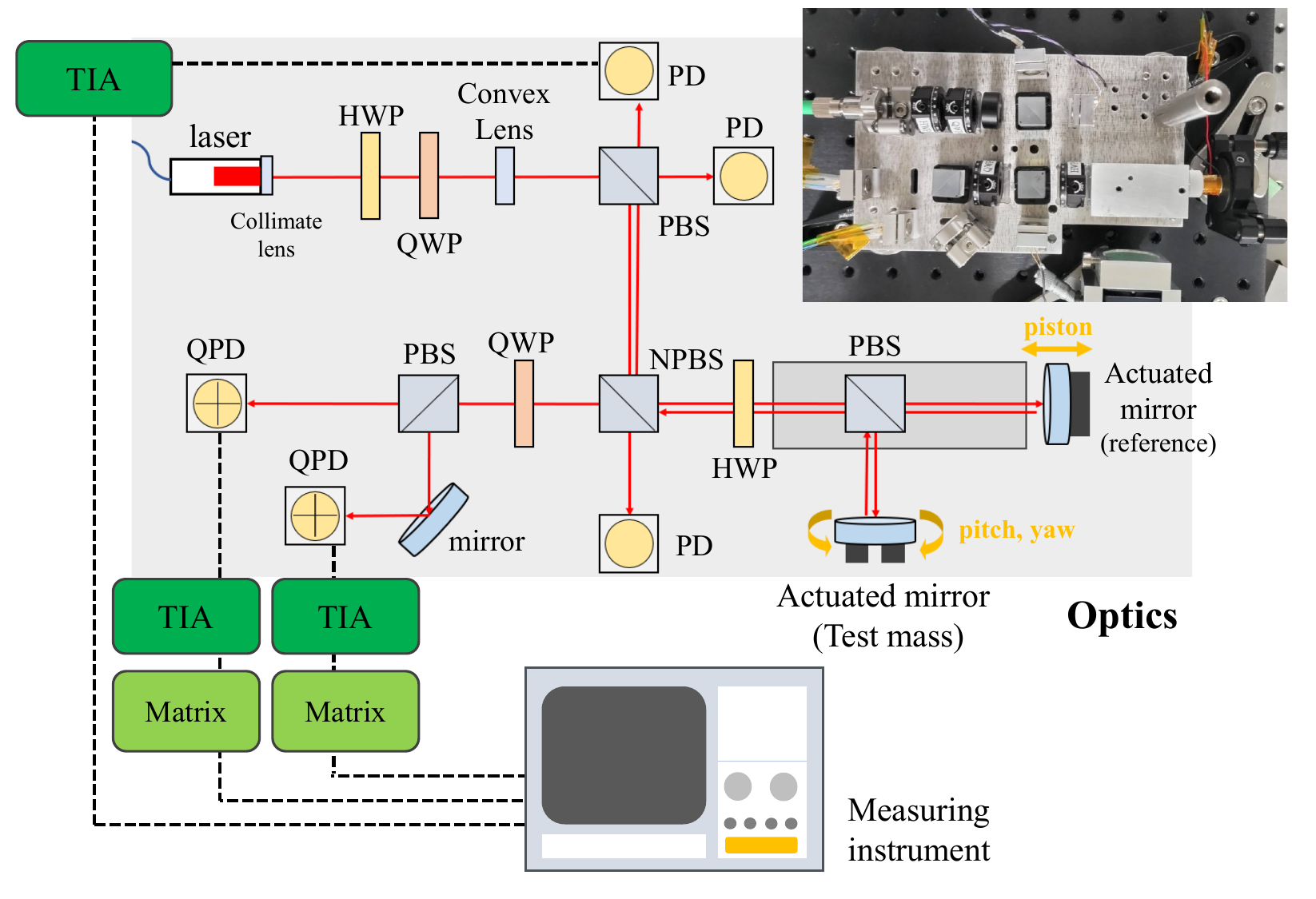}
  \caption{Optical layout used in the experiment. A photograph in the upper right corner shows the actual setup. The actuated mirror at the bottom simulates the tilt of the test mass while the one on the right of the PBS adjusts the interference condition. The yellow texts and accompanying arrows illustrate the direction of motion for the actuator coupled to the mirrors.}
  \label{fig:photo}
\end{figure}

In our formulation, the dependence of light intensities on the distance from the target mirror to the QPD and beam width at the QPD is implicitly included through the Gouy phase rather than explicitly expressed in terms of the geometric factor. 
This approach provides a more general description, particularly when the system size is comparable to or larger than the Rayleigh range.

\section{Experimental setup} \label{sec:es}
The experimental setup for demonstration of tilt measurement with QUIMETT is illustrated in Fig.~\ref{fig:photo}. 
In this experiment, a laser source, Mephisto S200 of Coherent Inc., was employed with the wavelength of 1064 nm. 
The free space output from the laser source is coupled to a single-mode optical fiber and introduced to the fiber collimator on the optical bench. 
The input power at NPBS is adjusted to be 1.6 mW.
The QPDs are G6849 from Hamamatsu Photonics. 
All the optical components except for two end mirrors are mounted on an aluminum plate. 
The mirror simulating the test mass is a broadband dielectric mirror with a diameter of 19~mm from Thorlabs, and the other optical elements are designed to have 1/2 inch in diameter. 

A convex lens is inserted in the optical system for two reasons. 
First, the lens keeps the beam size small enough throughout the system so that it does not require an additional train of lenses. 
The other reason is that it adjusts the Gouy phase evolution.
As discussed in Sec.~\ref{sec:wp}, the response to tilt deviations depends on the Gouy phase accumulated from the test mass to the locations of QPDs. 
If not carefully designed, the tilt signal would vanish in the worst case. 
We avoid such a situation by designing the beam waist to be located in the vicinity of the test mass. 
In our experiment, the beam waist position and the waist diameter are measured to be 15~mm behind the test mass and $0.2$ mm, respectively. 
The Gouy phase from the test mass to the QPDs is measured to be $124$ deg. 
Note that the beam diameter on the QPDs is $0.7$ mm.

All the QPDs and PDs are connected to the transimpedance amplifiers (TIAs) via lead wires. 
The TIAs are mounted on a separate location. 
The voltage signals coming out of the TIAs are continuously recorded in a time series. 
The measurement instrument is YOKOGAWA DL850E with 16-bit modules.

\section{Experimental Demonstrations}\label{sec:re}
\subsection{Tilt Measurement}
As discussed in Sec.~\ref{sec:wp}, the critical function in QUIMETT is the tilt sensing which is designed to be independent of the interference condition. 
To experimentally demonstrate this function, the tilt measurements were repeated over four different interference conditions in QPD1 as follows.
    \begin{equation}
        2k\left(l_x-l_y\right) = \begin{cases} \pi N &\textrm{(mid($-$))}\\
        \pi N + \pi/4 &\textrm{(dark)}\\
        \pi N + \pi/2 &\textrm{(mid($+$))}\\
        \pi N + 3\pi/4 &\textrm{(bright)}
        \end{cases} \label{eq:phiopt}
    \end{equation}
We hereafter call these interference conditions, mid($-$), dark, mid($+$), and bright. 
The interference condition was adjusted by the piezo actuator on the reference mirror. 

\begin{figure}[t]
  \includegraphics[width=\columnwidth]{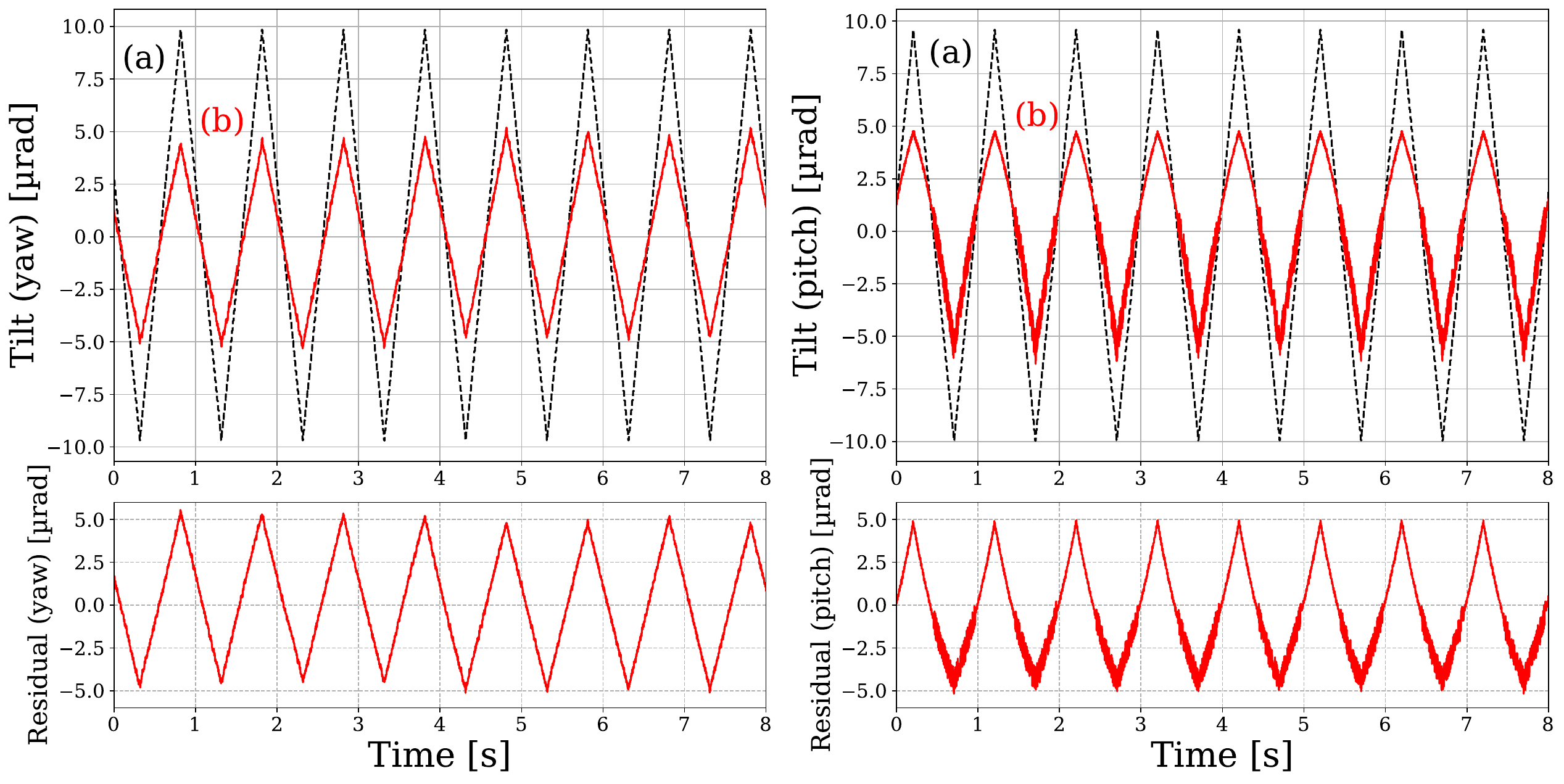} 
  \caption{Comparison of theoretical (a) and actual (b) tilt measurement results for yaw and pitch directions in bright fringe. The theoretical traces are based on Eq.~(\ref{eq:QUIMETTyaw}), described in Sec.~\ref{sec:wp}. The left figure shows the yaw measurements, and the right figure shows the pitch measurements.}
  \label{expectvsreal}
\end{figure}

During the measurements, the tilt of the test mass simulator was excited by triangular waves at 1~Hz with an amplitude of 10~$\mu$rad. 
The triangular waves were generated by a function generator and sent to the set of piezo actuators driving the test mass simulator as shown in Fig.~\ref{fig:photo}. 
The pitch and yaw excitations were applied one at a time.

The results of tilt measurements for yaw and pitch directions in the bright fringe are shown in Fig.~\ref{expectvsreal} together with the theoretical values for comparison. 
The measured voltage data are calibrated to the tilt data using Eq.~\eqref{eq:QUIMETTyaw}. 
In this calculation, we assume $R_\mathrm{PBS2} = 1$ and $T_\mathrm{PBS2} = 1$ and we use the parameters shown in Tab.~\ref{tab:params}. 
In both yaw and pitch directions, the measured values are found to be almost half of the theoretical expectation values. 
However, the tilt motions are measured consistently during the measurement time. 
Some discrepancy could be due to the uncharacterized optical losses in the optics in the actual setup. 
Further discussion is provided in Section~\ref{sec:dis}.
The tilt measurement and residual show the same waveform, indicating that there is not large distortion in the measurement but a difference in the scaling factor.

\begin{table}[htb]
    \centering
    \caption{Parameters used to calculate the theoretical output of the pitch and yaw signals shown in Figs.~\ref{expectvsreal} and \ref{fig:fringeeffect}.}
    \label{tab:params}
    \begin{tabular}{l|c}
        Parameters & Values  \\ \hline \hline
        Input power ($P_\mathrm{in}$) & 1.6 mW \\
        Gouy phase ($\eta_\mathrm{ITF} + \eta_\mathrm{QPD}$) & $124$ deg \\
        Response of QPDs & 0.68 A/W\\
        Trans-impedance gain of QPDs & 4700 V/A \\
        Beam diameter on the actuated mirror & 340 $\mu$m \\
        Beam diameter on the QPDs & 700 $\mu$m
    \end{tabular}
\end{table}

\begin{figure}[t]
\begin{minipage}[b]{\columnwidth}
  \centering
  \includegraphics[width=\columnwidth]{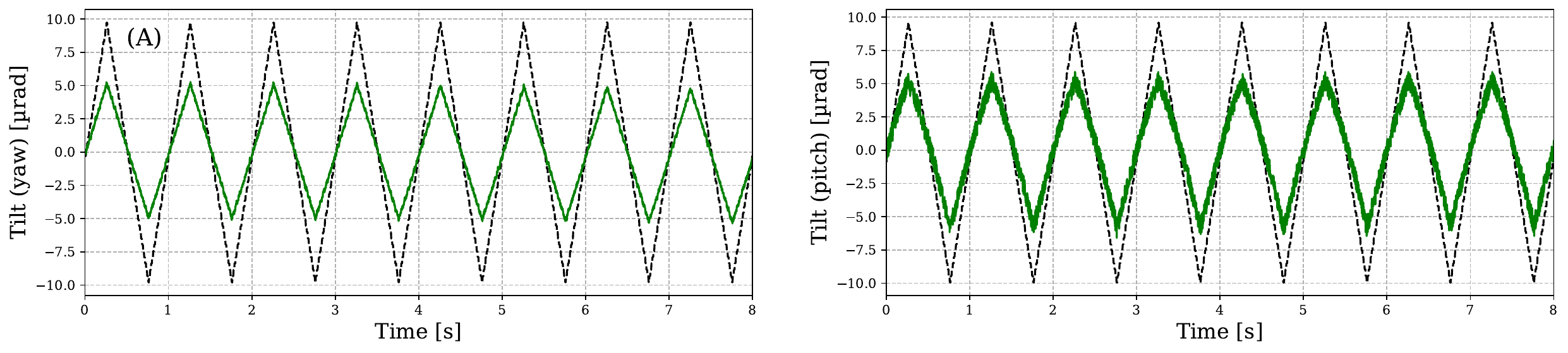}
\end{minipage}\\
\begin{minipage}[b]{\columnwidth}
    \centering
    \includegraphics[width=\columnwidth]{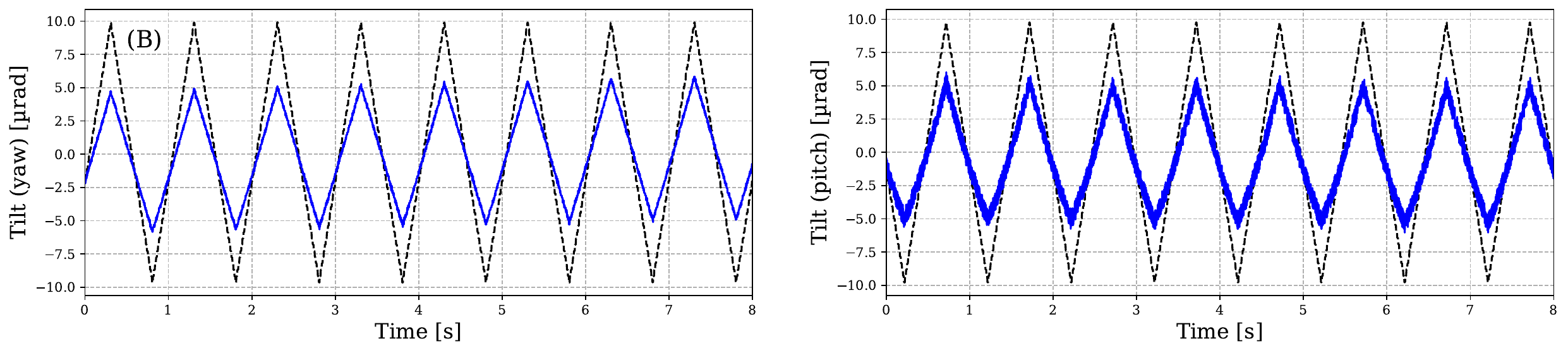}
    \end{minipage}\\
\begin{minipage}[b]{\columnwidth}
    \centering
    \includegraphics[width=\columnwidth]{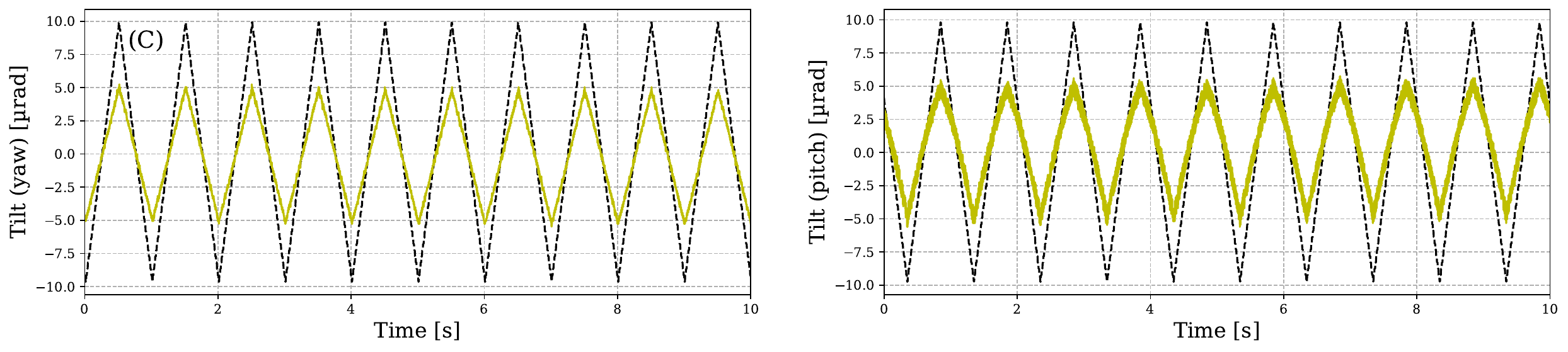}
\end{minipage}
\caption{Tilt measurements at different interference conditions. (A) shows mid($+$) fringe, (B) shows dark fringe, and (C) shows mid($-$) fringe.}
\label{fig:fringeeffect}
\end{figure}

Similarly, the results of the tilt measurements for the other three interference conditions are shown in Fig.~\ref{fig:fringeeffect}. 
It is clear that the tilt motion can be measured in three fringe conditions in addition to the mid($-$). 
Therefore, QUIMETT can read out the tilts without an influence from the interference condition as expected.

\subsection{Tilt measurement noise}
\begin{figure}
    \centering
    \includegraphics[scale=0.24]{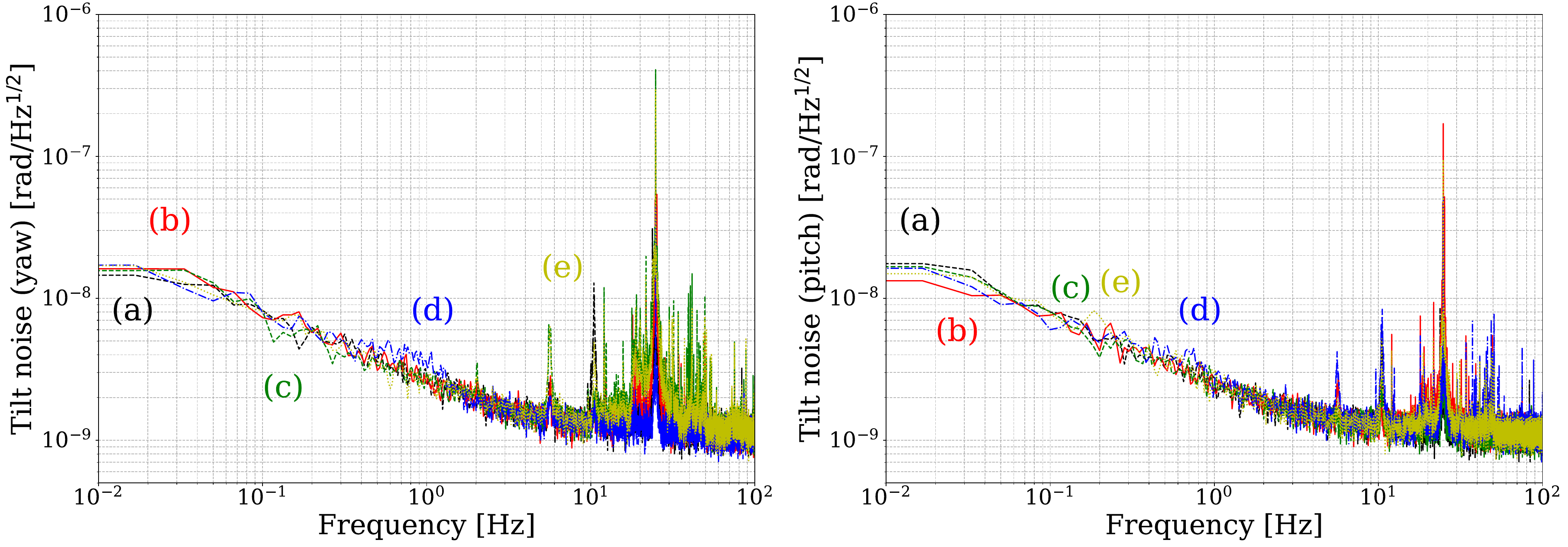}
    \caption{Tilt noises in amplitude spectral densities. Measurements were performed at four interference conditions. (b) shows bright fringe, (c) shows mid($+$), (d) shows dark fringe, and (e) shows mid($-$). The measurement results without the piezoelectric element (a) are also shown to evaluate the effect of the piezoelectric element.}
    \label{angularsensingnoise}
\end{figure}

The noise measurement was performed to assess the practical sensitivity with QUIMETT under the four interference conditions~(\ref{eq:phiopt}). 
The recorded time series were converted into the amplitude spectral density via the Welch's estimation technique with the Hann window applied. 
The frequency band was set to 0.01-100~Hz, which is relevant to the space gravitational wave antennas. 
The amplitude spectral densities of noise in four different interference conditions are shown in Fig.~\ref{angularsensingnoise}.

In addition, an extra measurement was performed with the mirror in the optical system replaced with a mirror mounted on a standard mount to see the influence from the piezo-actuated mirror. 
The data are shown with label (a) in Fig.~\ref{angularsensingnoise}.
 
\begin{figure}
    \centering
    \includegraphics[scale=0.3]{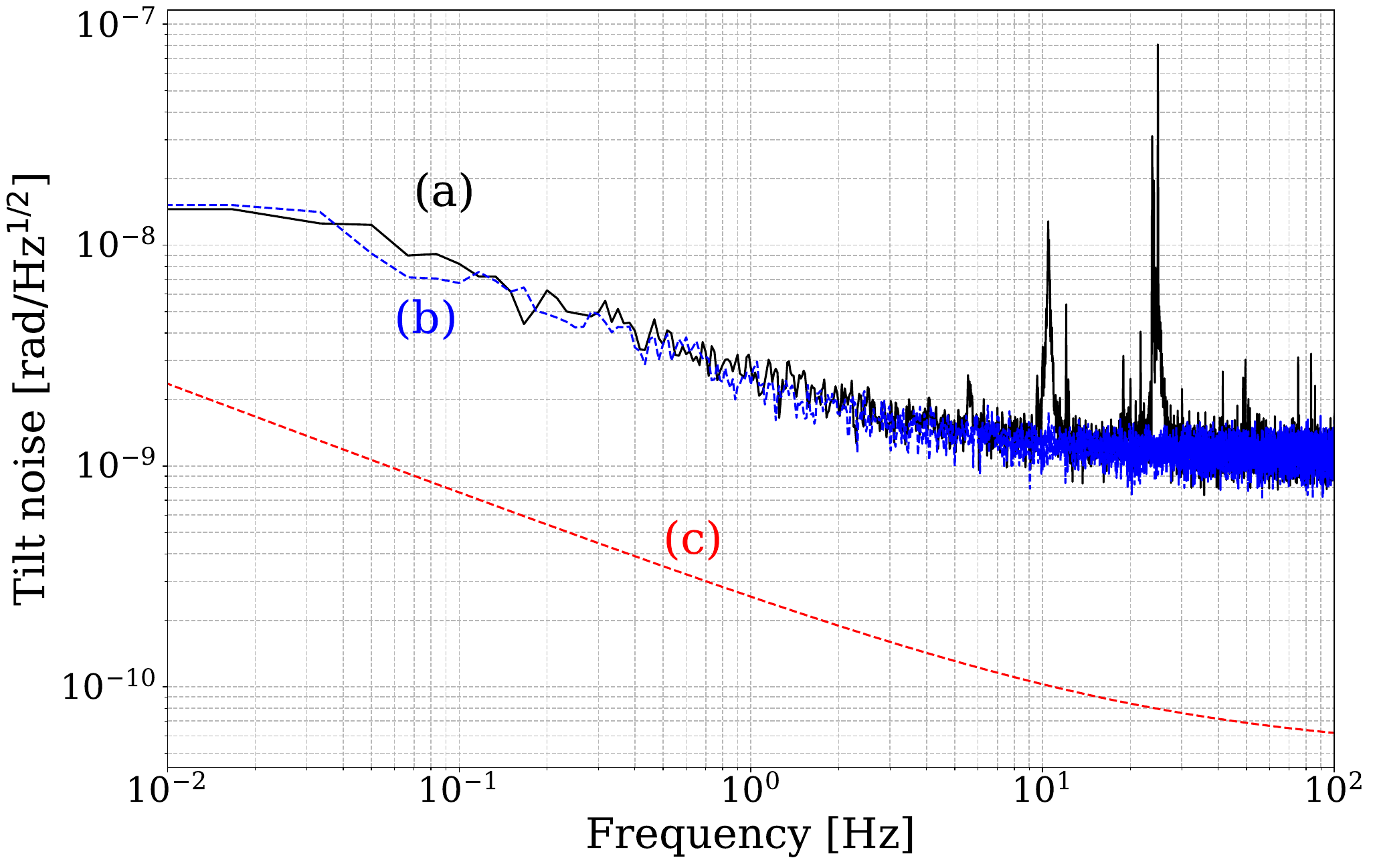}
    \caption{Noise budget for tilt noises. (a) shows the tilt sensing noise obtained with a fixed mirror, (b) shows the ADC noise associated with the measurement instrument and (c) shows electronic noise by the TIA and matrix electronics.}
    \label{noisebudget}
\end{figure}

In any interference conditions, sharp features are seen at frequencies above 10~Hz. 
These noise features increase in the mid fringes. 
On the other hand, we find the noise levels to be almost identical among the different interference conditions at frequencies below 10~Hz. 
We determined this noise to be the digitization noise from the data recorder as shown in Fig.~\ref{noisebudget}. 
In fact, the tilt sensing noise above 10~Hz is almost entirely dominated by the digitization noise.

\section{Discussions}\label{sec:dis}

Besides the demonstration of the tilt measurement, there are several critical considerations which have to be addressed before adopting QUIMETT as the local sensor for the drag-free control.

\subsection{Comparison between experimental and theoretical results}

Figures \ref{expectvsreal} and \ref{fig:fringeeffect} show that the theoretical predictions and experimental measurements exhibit a discrepancy of approximately a factor of two.
The tilt angle of the movable mirror has been independently verified using an optical lever within 1\% error. 
Therefore, we discuss systematic errors in other parts of the system as the possible source of the discrepancy.

Our analysis suggests that this discrepancy is not caused by a single major error. 
Instead, it results from the accumulation of several small systematic errors. 
These include minor deviations in mirror reflectivity, absorption losses in waveplates, imperfections in the extinction ratio of the PBS, variations in the quantum efficiency of the QPDs, and discrepancies in the transimpedance amplifier gain. 
Although each of these factors may have a limited individual impact, their combined effects contribute to the observed mismatch. 
Reducing these uncertainties is crucial for improving the accuracy of the system. Addressing these issues will be an important objective for future work.
Nevertheless, these systematic errors influence all fringe conditions in the same manner. 
As a result, the consistency of our experimental results across different fringe conditions suggests that their effects are uniform. 
This indicates that the proposed method successfully integrates tilt-sensing functionality into a quadrature interferometer. 
At the same time, it maintains the capability for displacement sensing over a wide range.

\subsection{Noise performance evaluation}

We also evaluated the noise performance of QUIMETT. 
Compared to conventional optical lever tilt sensors~\cite{Kokeyama:2018wpc}, our current setup exhibited a noise level approximately ten times higher. 
The primary reason for this increased noise is the lack of full optical optimization, particularly in terms of beam waist positioning and the collimator and lens configuration.

A major source of noise in the experiment was digitization noise from the ADC since, as shown in Eqs.~(\ref{eq:angQPD1}) and (\ref{eq:angQPD2}), each QPD signal contains not only the tilt component but also the one associated with the length difference or $l_x - l_y$. 
Consequently the electronics and measurement instrument have to deal with a wide range of signals. 
In the current setup, the tilt signals of each QPD were recorded by the measurement instrument and subsequently the sum of the two QPD signals was computed in the post process. 
This made the contribution of ADC noise particularly large due to the relatively narrow dynamic range in the ADC. 
Therefore, we can improve the noise floor by introducing analog electronics such that the summing process is completed before the ADC where the ADC range can be adjusted to be a smaller value.

\subsection{Coupling effects between translation and tilt}
A further look into the length-to-tilt coupling is necessary. 
As described in Sec.~\ref{sec:re}, the response to tilts did not change as a function of the interference condition. 
However, there could be unexpected cross-coupling where the dynamical motion in the length degree of freedom leaks into the tilt signals. 
We were not able to quantify such couplings in this setup because the piezo-actuated mirror showed an inherent cross-coupling between the tilts and length presumably due to the mechanical imbalance. 

Similarly, the tilt-to-length coupling must be studied. 
As shown in Eq.~(\ref{eq:dispreadout}), the length signal is not affected by variations in light intensity, as long as the efficiencies of the PD and QPDs remain constant. 
However, the total amount of light detected by the QPDs could easily be a function of the beam spot displacement caused by the test mass tilts because the QPDs are inherently more sensitive to this coupling due to the fact that there are gaps between the segments where the light is not converted into the photocurrent. 
Such a tilt-to-length coupling can also occur in the single-element PD due to the inhomogeneous responsivity across the PD surface area, but it is typically smaller than that of the QPDs.

Furthermore, comparing against the single-element PDs, the use of QPDs tends to reduce the signal-to-noise ratio for length sensing due to the presence of the gap between the segments. 
This could lead to a situation where the level of noise floor in the length measurement varies depending on the spot position on the QPDs. 
This may impact on the background noise estimation in gravitational wave data analysis where the stationary noise over a long time scale is often preferable.

\subsection{Optimization and future prospects}

For future applications of QUIMETT, several optimizations must be considered. 
Here, we consider the measurement range. 
The measurement range of our device is determined by the normalized tilt parameter $\Theta$ which was treated as a small second-order effect in Eqs.~(\ref{eq:IofQPD1}) and~(\ref{eq:IofQPD2}). 
This suggests that the current dynamic range is approximately $\sim 1$ mrad, where $\Theta^2$ remains on order of magnitude of unity. 
For future applications, such as the potential use of QUIMETT in DECIGO, we estimate the required measurement range by considering a test-mass module configuration similar to the LISA Pathfinder~\cite{LISAPathfinder:2023mgh} extended to DECIGO~\cite{nagano2020control}. 
Assuming a $4$ mm gap in the sensor module and a $30$ cm mirror size, the necessary range would be $\sim 10$ mrad. 
This exceeds the current capability, indicating that further design considerations, such as incorporating additional optical elements like lenses, will be necessary to extend the measurement range.

Finally, refinement of the optical design to reduce the volume and weight for the space application is necessary. 
Equation~(\ref{eq:QUIMETTyaw}) reveals that optical path length does not influence the tilt sensitivity. 
This means the reduction of the optical path lengths enables a design with smaller volume without compromising the sensitivity. 
In particular, shortening the distance between PBS1 and NPBS in Fig.~\ref{fig:concept} is considered to be effective for achieving a compact design.

\section{Conclusion}\label{sec:con}
This paper proposes a new local sensor concept for the drag-free control in space. 
The sensor is based on the quadrature interferometer and named QUIMETT. 
QUIMETT enables the simultaneous readout of both translation and tilt degrees of freedom using a single interferometer module. 
This paper presents for the first time the working principle of the tilt measurements and its demonstration by building the dedicated optical system. 
In the experimental verification, the main focus was put on the demonstration of the tilt readout which was designed to be independent of the interference condition. 
The experimental demonstration successfully confirmed this function. 
The tilt sensitivity currently marks $10$~nrad/Hz$^{1/2}$ at $0.1$~Hz and did not change in four different interference conditions as expected. 
We discussed the studies which have to be further looked in before considering this scheme as a viable solution for the drag-free control.

\section*{Acknowledgment}
We thank Advanced Machining Technology Group and K. Komori for their technical support. 
We also appreciate C. M. Mow-Lowry and S. Sato for their useful discussions.
This project was supported by JSPS KAKENHI Grant Number JP20J01928 and JP22K14067. K.N. is an employee of LQUOM, Inc.

\section*{Author contributions}
K.N. and K.M. carried out the experiment.
K.N. conceived the original idea.
K.M. and K.I. developed the theoretical formalism, and performed the analytic calculations.
K.I. supervised the project.
All authors discussed the results and contributed to the final manuscript.
K.N. and K.M. contributed equally to the present work.

\appendix
\section{Mathematical foundation}
\label{AppendixA}

Here, we show the mathematical background of the optical analysis in this paper. 
Upon transmission and reflection, the electric field in a polarization acquires sets of coefficients given matrix form as,
\begin{equation}
\hat{t}\left(T\right)=\sqrt{T}\,
\textrm{diag}\left(1,1,1\right)
\quad\textrm{and}\quad    \hat{r}\left(R\right)=\sqrt{R}\,
\textrm{diag}\left(1,-1,1\right),
\end{equation}
where diag represents the diagonal matrix, $T$ and $R$ are the transmittance and reflectance of the optical component, respectively.
Using these matrices, one can express the polarizing beam splitter (PBS) in the $6\times 6$ form as,
\begin{equation}
    \mathbb{B}_h\left(T\right)=\begin{bmatrix}
        \hat{t}(T)&\hat{0}\\\hat{0}&\hat{0}
    \end{bmatrix}\quad\textrm{and}\quad\mathbb{B}_v(R)=\begin{bmatrix}
        \hat{0}&\hat{0}\\\hat{0}&\hat{r}(R)
    \end{bmatrix}.
    \label{eq:PBS}
\end{equation}
$\mathbb{B}_t$ is transmission of PBS, and $\mathbb{B}_v$ is reflection of PBS.
Similarly, reflection and transmission of optical elements are expressed as
\begin{equation}
    \mathbb{A}_\textrm{t}(T)=\begin{bmatrix}
        \hat{t}(T)&\hat 0\\\hat{0}&\hat{t}(T)
    \end{bmatrix}\quad\textrm{and}\quad\mathbb{A}_\textrm{r}(R)=  \begin{bmatrix}
        -\hat{r}\left(R\right) &\hat 0\\\hat{0}&\hat{r}\left(R\right)
    \end{bmatrix}.
\end{equation}
We deal with misalignment of an optical element~\cite{hefetz} as an additive matrix as
\begin{equation}
\hat{M}\left(\Theta_\textrm{yaw},\Theta_\textrm{pitch}\right)=
    \begin{bmatrix}
        0&-2i\Theta_\textrm{yaw}&-2i\Theta_\textrm{pitch}\\-2i\Theta_\textrm{yaw}&0&0\\-2i\Theta_\textrm{pitch}&0&0
    \end{bmatrix} ,
    \label{mat:misalign}
\end{equation}
where $\Theta_A$ represents the normalized tilt. 
It is defined by $\Theta_A=(\theta_A\pi\omega(z))/\lambda$, where
$\theta_A$ is a rotation in the optical element, $\lambda$ is the wavelength and $w(z)$ is the beam radius at the propagation point $z$. 
Equation~(\ref{mat:misalign}) is only valid when $\Theta_\textrm{A}$ is sufficiently small.
The propagation through an optical path transmission is represented as
\begin{equation}
\mathbb{P}\left(l,\eta\right)=e^{-ikl}\mathbb{R}(\eta),
    \label{mat:pathx}
\end{equation}
where $l$ is the distance from one location to another, $\eta$ is the Gouy phase shift in the optical path, and $k$ is the wave number. 
The Gouy phase shift matrix $\mathbb{R}(\eta)$ is expressed as
\begin{align}
    \mathbb{R}(\eta)=\begin{bmatrix}
    \hat{R}(\eta)&\hat{0}\\\hat{0}&\hat{R}(\eta)
\end{bmatrix}\quad\textrm{where}\quad\hat{R}(\eta)=\textrm{diag}\left(1,e^{-i\eta},e^{-i\eta}\right) .
\label{eq:gouyphase}
\end{align}
The half-wave plate is expressed as~\cite{jones}
\begin{equation}
    \mathbb{W}^{(2)}\left(\phi\right)=\begin{bmatrix}
        \hat{H}\left(\phi\right)&\hat{H}\left(\frac{\pi}{4}-\phi\right)\\\hat{H}\left(\frac{\pi}{4}-\phi\right)&-\hat{H}\left(\phi\right)
    \end{bmatrix},
\quad \textrm{where}\quad
    \hat{H}\left(\phi\right)=\cos{2\phi}
   \hat{I},
    \label{mat:HWP}
\end{equation}
where $\phi$ is the fast-axis angle of the half-wave plate. 
Jones matrix of quarter-wave plate is expressed as,
\begin{equation}
\begin{split}
    \mathbb{W}^{(4)}\left(\psi\right)=\begin{bmatrix}
        \frac{1+i}{2}\hat{I}+\hat{Q}\left(\frac{\pi}{4}-\psi\right)&\hat{Q}\left(\psi\right)\\\hat{Q}\left(\psi\right)&\frac{1+i}{2}\hat{I}-\hat{Q}\left(\frac{\pi}{4}-\psi\right)
    \end{bmatrix},\quad \textrm{where}\quad\hat{Q}\left(\psi\right)=\frac{1-i}{2}\sin{2\psi}\hat{I} ,
    \label{mat:QWP}
    \end{split}
    \end{equation}
where $\psi$ is the fast-axis angle of the quarter-wave plate. 

The total intensity of laser falling onto a PD or QPDs can be calculated as
\begin{equation}
    I^\textrm{sum}= \sum_{A=h,v} \left(\vec{E^A}\right)^\dagger \vec{E^A}.
    \label{eq:howtoPDsum}
\end{equation}
For tilt sensing by the QPDs, we consider the laser intensity in the hemisphere of the QPD. 
To detect tilt in the yaw direction, the laser intensity in the left and right halves is discussed here. 
We hereafter consider one polarization mode since, in the setup of QUIMETT, the QPDs detect one polarized light due to the PBS in front of the QPDs. 
The laser intensity of the right half of the QPD is denoted as
\begin{align}
    S^\textrm{(r.h.)}&=\int^\infty_0|\vec{E}|^2dx=\int^\infty_0|E_{00}U_{00}+E_{10}U_{10}+E_{01}U_{01}|^2dx
    \\&=\frac{1}{2}|E_{00}|^2 +\left[E_{00}E_{10}^{*}\int^\infty_0 U_{00}U^{*}_{10}dx+(\textrm{c.c})\right] + (\textrm{vertical mode term}),
    \label{eq:srh1}
    \end{align}
where $x$ represents horizontal position on the QPD, and $U_{00}$ and $U_{10}$ ($U_{01}$) represent the fundamental and first-order Hermite polynomials denoting the horizontally (vertically) spatial mode of the Gaussian-Hermite beam, respectively. 
Here, we assume that $|E_{00}| \gg |E_{10/01}|$ and the size of the QPD is much larger than the beam size on the QPD. 
$U_{00}$ and $U_{10}$ are defined as
\begin{equation}
    U_{00}=\left(\frac{2}{\pi}\right)^\frac{1}{4}\frac{1}{\sqrt{\omega}}\exp{\left[-x^2\left(\frac{1}{\omega^2}+i\frac{k}{2R}\right)\right]}
    \quad\textrm{and}\quad U_{10}=\frac{2x}{\omega}U_{00},
    \label{eq:u0u1}
    \end{equation}
where $\omega$ and $R$ are the beam size and the curvature of the electric field on the QPD, respectively. 
From this definition, we can obtain the following equation, 
\begin{align}
    \int^\infty_0 U_{00} U_{10}^*dx&=\left(\frac{2}{\pi}\right)^{\frac{1}{2}}\frac{2}{\omega^2}\int^\infty_0x\exp\left(-\frac{2x^2}{\omega^2}\right)dx\notag
    \\&=\frac{1}{\sqrt{2\pi}}.
\end{align}
Also this is true for the complex conjugated one, so that
\begin{align}
    \int^\infty_0U_{00}U_{10}^*dx=\int^\infty_0 U_{00}^*U_{10}dx=\frac{1}{\sqrt{2\pi}}.
\end{align}
As a result, Eq.~\eqref{eq:srh1} is denoted as
\begin{align}
    S^\textrm{(r.h.)}=\frac{1}{2}|E_{00}|^2+\sqrt{\frac{2}{\pi}}\Re[E_{00}E_{10}^*] + (\textrm{vertical mode term}).
    \label{eq:half}
\end{align}
Similarly, the left half is:
\begin{align}
    S^\textrm{(l.h.)}=\frac{1}{2}|E_{00}|^2-\sqrt{\frac{2}{\pi}}\Re[E_{00}E_{10}^*] + (\textrm{vertical mode term}).
    \label{eq:otherhalf}
\end{align}
Tilt sensing by the QPD is provided by subtraction of one hemisphere from the other,
\begin{align}
    S^\textrm{yaw}=S^\textrm{(r.h)}-S^\textrm{(l.h)}=\sqrt{\frac{8}{\pi}}\Re[E_{00}E_{10}^*].
    \label{eq:tiltpfQPD}
\end{align}
Substituting $E_{01}^*$ (vertical mode) for $E_{10}^*$ (horizontal mode) in Eq.~(\ref{eq:tiltpfQPD}) allows us to derive the pitch tilt detection value.

\section{Effect of reference mass}
\label{AppendixB}

Here, we study the case of tilting $\theta_\textrm{yaw2}$, and $\theta_\textrm{pitch2}$ of the reference mirror. 
Under this condition, the electric field incident upon the two QPDs can be derived as
\begin{equation}
\begin{split}
 \mathbb{E}^{\textrm{QPD}n}=&\mathbb{F}_n\mathbb{W}^{\left(4\right)}\left(\frac{\pi}{4}\right)\mathbb{P}(l_3,\eta_{3})\mathbb{A}_\textrm{t}\left(\frac{1}{2}\right)\mathbb{P}(l_1,\eta_{1})\mathbb{W}^{\left(2\right)}\left(-\frac{\pi}{8}\right)\\&\times\lbrace\mathbb{B}_v\left(R_\textrm{PBS2}\right)\mathbb{P}(l_y,\eta_\mathrm{ITF})\mathbb{M}\left(\Theta_\textrm{yaw},\Theta_\textrm{pitch}\right)\mathbb{P}(l_y,\eta_\mathrm{ITF_{in}})\mathbb{B}_v\left(R_\textrm{PBS2}\right)\\&+\mathbb{B}_h(T_\textrm{PBS2})\mathbb{P}(l_x,\eta_\mathrm{ITF})\mathbb{M}\left(\Theta_\textrm{yaw2},\Theta_\textrm{pitch2}\right)\mathbb{P}(l_x,\eta_\mathrm{ITF_{in}})\mathbb{B}_h\left(T_\textrm{PBS2}\right)\rbrace \\& \times \mathbb{W}^{\left(2\right)}\left(\frac{\pi}{8}\right)\mathbb{P}(l_1,\eta_{1_\mathrm{in}})\mathbb{A}_\textrm{r}\left(\frac{1}{2}\right)\mathbb{E}_\textrm{in} . \label{eq:EQPDref}
 \end{split}
\end{equation}
The tilt signal of QUIMETT is
\begin{align}
    I^\textrm{tilt}_\textrm{QPD1}+I^\textrm{tilt}_\textrm{QPD2} = \frac{\left(\Theta_{B\textrm{2}}T_\textrm{PBS2}^2 + \Theta_{B}R_\textrm{PBS2}^2\right)\sin{\left(\eta_\mathrm{ITF}+\eta_\mathrm{QPD} \right)}}{\sqrt{2\pi}}P_\textrm{in} .
    \label{eq:QUIMETTref}
\end{align}
Equation (\ref{eq:QUIMETTref}) shows the tilts of the two mirrors are inseparable. 
Therefore, it is important to mount the reference mirror securely for precise tilt measurement.

\section*{References}
\bibliographystyle{unsrt}
\bibliography{ref}

\end{document}